\documentstyle[preprint,epsf,aps]{revtex}  
\begin{document}
\draft         

\preprint{DFNAE-IF/UERJ-98/04, IFUSP/P-1312}

\title{Pair Production of Heavy-Exotic-Fermions}

\author{ J.\ E.\ Cieza Montalvo $^*$} 

\address{Instituto de F\'{\i}sica, Universidade do Estado do Rio de Janeiro\\ 
CEP $20559-900$ Rio de Janeiro, Brazil}

\address{and Instituto de F\'{\i}sica, Universidade de S\~ao Paulo \\
C.P.\ 66.318, 05315--970, S\~ao Paulo -- SP, Brazil}


\maketitle


\begin{abstract} 

We study the production and signatures of heavy exotic fermion pairs
in the framework of the vector singlet model (VSM), vector doublet
model (VDM) and fermion-mirror-fermion (FMF) model. We show that the
pair production cross sections are competitive with the ones for the
single production of exotic fermions due to the exchange of a photon
in the $s$ channel and to the vector component of the $Z$ - boson
coupling to exotic fermions. We also exhibit some kinematical
distributions.


\vskip 2.5cm
{\em Submitted to Phys.\ Rev.\ D}
\end{abstract}

\newpage

\section{INTRODUCTION}

The standard electroweak theory provides a very satisfactory description of
most elementary particle phenomena up to the presently available
energies. However, there are experimental facts, such as the proliferation of
the fermion generation and their complex pattern of masses and mixing angles,
that are not predicted in the framework of the standard model. There is no
theoretical explanation for the existence of several generations and for the
values of the masses. It was established at the CERN $e^{+} e^{-}$ collider
LEP that the number of light neutrinos is three \cite{lep}. However this
result does not imply that further generations cannot exist
\cite{nov}.

Many models, such as composite models \cite{af,bu1}, grand unified theories
\cite{la}, technicolor models \cite{di}, superstring-inspired models
\cite{e6}, mirror fermions \cite{maa}, predict the existence of new particles
with masses in turn of the scale of $1$ TeV and they consider the possible
existence of a new generation of fermions.

In this paper, we will study the production mechanism for exotic-fermions in
$e^{+} e^{-}$ colliders such as NLC ($\sqrt{s} = 500$ GeV) and CLIC ($\sqrt{s}
= 1000$ GeV). It is assumed here that the mixing between ordinary and exotic
leptons are of the same flavour. We also note that the electromagnetic
current do not present any alteration in comparison with the standard model, when mixing between ordinary and exotic particles are of equal charges \cite{sim4,sim3}.

The outline of this paper is the following. The models and mixing angles are presented in Sec. II. In Sec. III, we study the pair production of exotic-fermions. We summarize our results in Sec. IV


\section{MODELS AND MIXING ANGLES}

The models that we consider in this work include new
fermionic degrees of freedom, which introduce naturally a number of
unknown mixing angles and fermionic masses \cite{tom}.

We will consider for the production of exotic-fermions three
models: the vector singlet model (VSM) \cite{gon}, which consists
in the inclusion of new left- and right-handed fermions in
singlets, 
\[
\left(
\begin{array}{cc}
\nu&\\
e&\\
\end{array}
\right)_{L}, e_{R}, \nu_{R}, E_{L}, E_{R}, N_{L}, N_{R}, U_{L}, U_{R}, D_{L}, D_{R}; 
\]
the vector doublet model (VDM) \cite{riz}, that can arise at low energies in the ${\bf 27}$ representation of $E_{6}$ theories and it  includes new currents of the type:

\[
\left(
\begin{array}{cc}
N&\\
E&\\
\end{array}
\right)_{L}, \hskip 0.6cm
\left(
\begin{array}{cc}
N&\\
E&\\
\end{array} 
\right)_{R}, \hskip0.6cm 
\left(
\begin{array}{cc}
U&\\
D&\\
\end{array}
\right)_{L}, \hskip 0.6cm
\left(
\begin{array}{cc}
U&\\
D&\\
\end{array} 
\right)_{R} 
\]
and the fermion-mirror-fermion (FMF) model
\cite{maa}, where the new particles are introduced to restore the
right-left symmetry :

\[
\left(
\begin{array}{cc}
N&\\
E&\\
\end{array}
\right)_{R}, \hskip 0.6cm
\left(
\begin{array}{cc}
N&\\
\end{array} 
\right)_{L}, \hskip 0.6cm
\left(
\begin{array}{cc}
E&\\
\end{array}
\right)_{L}, \hskip 0.6cm 
\left(
\begin{array}{cc}
U&\\
D&\\
\end{array}
\right)_{R}, \hskip 0.6cm
\left(
\begin{array}{cc}
U&\\
\end{array} 
\right)_{L}, \hskip 0.6cm
\left(
\begin{array}{cc}
D&\\
\end{array}
\right)_{L} .
\]

Exotic fermions mixed with the standard fermions interact through the
standard weak vector bosons $W^{+}, W^{-}$ and $Z^{0}$, according to the Lagrangians

\begin{equation}
{\cal L}_{\rm NC} = \frac{g}{4 cos\theta_{W}} \left[{ \bar{F}_{i}}
\gamma^{\mu} (g_{V}^{ij} - g_{A}^{ij} \gamma^{5}) F_{j} +
{\bar{F}_{i}} \gamma^{\mu} (g_{V}^{ij} - g_{A}^{ij} \gamma^{5}) f_{j} \right]
Z_{\mu} 
\label{lag1}
\end{equation}
   
and 

\begin{equation}
{\cal L}_{\rm CC} = \frac{g}{2 \sqrt{2}} { \bar{L}^{0} \gamma^{\mu} 
(C_{V}^{ij} - C_{A}^{ij} \gamma^{5}) e^{-} W_{\mu} } ,
\label{lag2}
\end{equation}
where $g_{V}^{ij}$ and $g_{A}^{ij}$ are the corresponding neutral
vector-axial coupling constants and $C_{V}^{ij}$ and $C_{A}^{ij}$ are
the charged vector-axial coupling constants, which are given in Table I
\cite{sim4,sim3}, for each of the three models that we study here.

Since we have considered heavy fermion production mediated by the
$\gamma$, $Z$ and $W$, then, consequently, there appear a
neutral-current coupling and a charged-current coupling. For the
second case, we have for the VSM a $V-A$ coupling. {}For the VDM,  we have $V + A$ and for the FMF model, an axial vector (if all angles are equal), while for the first case we have neither of the possibilities $V -A$, $V + A$, V, nor A, since there appear a $\sin^{2} \theta_{W}$
\cite{sim4}.

We consider here that all mixing angles have the value $\theta_{i}
= 0.1$, although phenomenological analysis \cite{tom1} give an upper
bound of $sin^{2} \theta_{i} \leq 0.03$. This means that the value of
$\theta_{i}$ can be scaled up to $0.173$.

\hskip 0.5cm

TABLE I. Coupling constants for a charged and neutral heavy fermion
interaction: for the vector singlet model (VSM), the vector doublet model (VDM) and the fermion-mirror-fermion (FMF) model :

\vskip 1cm

\begin{tabular}{|l|r|c|r|}  \hline\hline
Cou. & VSM & VDM & FMF \\ \hline $g_{V}^{LL}$ & $-\frac{1}{2} \sin^{2}
\theta_{L}^{e} + 2 \sin^{2} \theta_{W}$ & $-\frac{1}{2} - \frac{1}{2}
\cos^{2} \theta_{R}^{e} + 2 \sin^{2} \theta_{W}$ & $-\frac{1}{2}
\sin^{2} \theta_{L}^{e} + \frac{1}{2} \cos^{2} \theta_{R}^{e} + 2
\sin^{2} \theta_{W}$ \\ \hline $g_{A}^{LL}$ & $-\frac{1}{2} \sin^{2}
\theta_{L}^{e}$ & $-\frac{1}{2} \sin^{2} \theta_{R}^{e}$ &
$-\frac{1}{2} \sin^{2} \theta_{R}^{e} - \frac{1}{2} \cos^{2}
\theta_{R}^{e}$ \\ \hline $g_{V}^{eL}$ & $-\frac{1}{2}\sin
\theta_{L}^{e} \cos \theta_{L}^{e}$ & $\frac{1}{2} \cos \theta_{R}^{e}
\sin \theta_{R}^{e}$ & $\frac{1}{2} \sin (\theta_{R}^{e} -
\theta_{L}^{e})$ \\ \hline $g_{A}^{eL}$ & $-1/2 \sin \theta_{L}^{e}
\cos \theta_{L}^{e}$ & $-\frac{1}{2}\sin \theta_{R}^{e} \cos
\theta_{R}^{e}$ & $-\frac{1}{2} \sin (\theta_{R}^{e} +
\theta_{L}^{e})$ \\ \hline $C_{V}^{eL^{0}}$ & $\sin \theta_{L}^{\nu}
\cos \theta_{L}^{e}$ & $\sin (\theta_{L}^{\nu} - \theta_{L}^{e}) -
\cos \theta_{R}^{\nu} \sin \theta_{R}^{e}$ & $\sin \theta_{L}^{\nu}
\cos \theta_{L}^{e} - \sin \theta_{R}^{e} \cos \theta_{R}^{\nu}$ \\
\hline $C_{A}^{eL^{0}}$ & $\sin \theta_{L}^{\nu} \cos \theta_{L}^{e}$
& $\sin (\theta_{L}^{\nu} - \theta_{L}^{e}) + \cos \theta_{R}^{\nu}
\sin \theta_{R}^{e}$ & $\sin \theta_{L}^{\nu} \cos \theta_{L}^{e} +
\sin \theta_{R}^{e} \cos \theta_{R}^{\nu}$ \\ \hline $g_{V}^{UU}$ &
$\frac{1}{2} \sin^{2} \theta_{L}^{u} -
\frac{4}{3} \sin^{2} \theta_{W}$  &  $\frac{1}{2} + \frac{1}{2} \cos^{2} \theta_{R}^{u} - \frac{4}{3} \sin^{2}  \theta_{W}$  &  $\frac{1}{2} \sin^{2} \theta_{L}^{u} + \frac{1}{2} \cos^{2} \theta_{R}^{u} - \frac{4}{3} \sin^{2} \theta_{W}$  \\   \hline 
$g_{A}^{UU}$  &  $\frac{1}{2} \sin^{2} \theta_{L}^{u}$  &  $\frac{1}{2} \sin^{2} \theta_{R}^{u}$  &   $\frac{1}{2} \sin^{2} \theta_{L}^{u} - \frac{1}{2} \cos^{2} \theta_{R}^{u}$  \\  \hline
$g_{V}^{DD}$  &  $-\frac{1}{2} \sin^{2} \theta_{L}^{d} - 
\frac{2}{3} \sin^{2} \theta_{W}$  &  $-\frac{1}{2} - \frac{1}{2} \cos^{2} \theta_{R}^{d} + \frac{2}{3} \sin^{2}  \theta_{W}$  &  $-\frac{1}{2} \sin^{2} \theta_{L}^{d} - \frac{1}{2} \cos^{2} \theta_{R}^{d} + \frac{2}{3} \sin^{2} \theta_{W}$  \\   \hline 
$g_{A}^{DD}$  &  $-\frac{1}{2} \sin^{2} \theta_{L}^{d}$  &  $-\frac{1}{2} \sin^{2} \theta_{R}^{d}$  &   $-\frac{1}{2} \sin^{2} \theta_{L}^{d} + \frac{1}{2} \cos^{2} \theta_{R}^{d}$  \\  \hline
\end{tabular}


\section{CROSS SECTION PRODUCTION}

Pair production of exotic particles is, to a very good
approximation, a model independent process, since it proceeds through a well known electroweak interaction. Then this production mechanism can be studied through the analysis of the reactions $e^{+} e^{-} \rightarrow F^{+} F^{-}$ and $e^{+} e^{-} \rightarrow F^{0} F^{0}$, provided that there is enough available energy ($\sqrt{s} \geq 2M_{L}$). These processes take place through the exchange of a photon in the $s$ channel, a boson $Z^{0}$ in the $s$ and $t$ channel and a boson $W$ in the $t$ channel. Using the interactions Lagrangians, Eqs. ($1$) and ($2$),  it is easy to evaluate the cross section for the process $e^{+} e^{-} \rightarrow L^{+} L^{-}$, involving a neutral current, from which we obtain:

\hskip 0.5cm

\begin{eqnarray} 
\left (\frac{d \sigma}{d\cos \theta} \right )_{L^{+} L^{-}} = &&\frac{\beta \alpha^{2} \pi}{s^{2}} \Biggl [\frac{1}{s} ( 2 s M_{L}^{2} + (M_{L}^{2} - t)^{2} + (M_{L}^{2} - u)^{2}  )  \nonumber \\
&&+ \frac{1}{2 \sin^{2} \theta_{W} \cos^{2} \theta_{W} (s - M_{Z}^{2})} ( 2 s M_{L}^{2} g_{V}^{LL} g_{V}^{l}  \nonumber \\
&&+g_{V}^{LL} g_{V}^{l} ( (M_{L}^{2} - t)^{2} + (M_{L}^{2} - u)^{2}  )  + g_{A}^{LL} g_{A}^{l} ( (M_{L}^{2} - u)^{2} - (M_{L}^{2} - t)^{2} )   \nonumber \\
&&- \frac{1}{2 \sin^{2} \theta_{W} \cos^{2} \theta_{W} (t - M^{2}_{Z})} (g_{V}^{{eL}^{2}} + g_{A}^{{eL}^{2}} ) ( s M^{2}_{L} + 
(M^{2}_{L} - u)^{2} ) \Biggr ]   \nonumber \\
&&+ \frac{\beta G_{F}^{2} M_{Z}^{4}}{8 \pi s} \frac{1}{(s - M_{Z}^{2} )^{2}} \Biggl [(g_{V}^{{LL}^{2}} + g_{A}^{{LL}^{2}}) (g_{V}^{{l}^{2}} + g_{A}^{{l}^{2}}) ((M_{L}^{2} - u)^{2}  \nonumber \\
&&+(M_{L}^{2} - t)^{2}) + 2 (g_{V}^{{LL}^{2}} - g_{A}^{{LL}^{2}}) (g_{V}^{{l}^{2}} + g_{A}^{{l}^{2}}) s M_{L}^{2}   \nonumber \\
&&+ 4 g_{V}^{LL} g_{A}^{LL} g_{V}^{l} g_{A}^{l} ((M_{L}^{2} - u)^{2} - (M_{L}^{2} - t)^{2}) \bigr )    
+ \frac{1}{ (t - M_{Z}^{2} )^{2}} \bigl ((g_{V}^{{eL}^{2}}  \nonumber \\ &&-g_{A}^{{eL}^{2}})^{2} \frac{1}{2} s^{2} (1 + \beta^{2}) + 
(g_{V}^{{eL}^{2}} + g_{A}^{{eL}^{2}})^{2}  (M_{L}^{2} - u)^{2}  
\nonumber \\
&&+4 g_{V}^{{eL}^{2}} g_{A}^{{eL}^{2}} (M_{L}^{2} - u)^{2} \bigr )  
- \frac{2}{s - M_{Z}^{2}} \frac{1}{t - M_{Z}^{2}} \bigl ((g_{V}^{LL} g_{V}^{{eL}^{2}} g_{V}^{l} + 2 g_{V}^{LL} g_{V}^{eL} g_{A}^{eL}
g_{A}^{l}   \nonumber \\ 
&&+g_{V}^{LL} g_{A}^{{eL}^{2}} g_{V}^{l} - g_{A}^{LL} g_{V}^{{eL}^{2}} g_{A}^{l} 
- 2 g_{A}^{LL} g_{V}^{eL} g_{A}^{eL} g_{V}^{l} -  g_{A}^{LL} g_{A}^{{eL}^{2}} g_{A}^{l}) s M_{L}^{2}  \nonumber  \\
&&+ (g_{V}^{LL} g_{V}^{{eL}^{2}} g_{V}^{l} + 2 g_{V}^{LL} g_{V}^{eL} g_{A}^{eL} g_{A}^{l}  
+g_{V}^{LL} g_{A}^{{eL}^{2}} g_{V}^{l}  
+g_{A}^{LL} g_{V}^{{eL}^{2}} g_{A}^{l}   \nonumber  \\
&&+ 2 g_{A}^{LL} g_{V}^{eL} g_{A}^{eL} g_{V}^{l} +  g_{A}^{LL} g_{A}^{{eL}^{2}} g_{A}^{l}) 
(M_{L}^{2} - u)^{2}  ) \Biggr ]      \; ,
\end{eqnarray}
where $\beta_{L} = \sqrt{1- 4 M_{L}^{2}/s}$ is the velocity of
exotic-fermion in the c.m. of the process, $G_{F}$ is the Fermi
coupling constant, $M_{Z}$ is the mass of the $Z$ boson, $\sqrt{s}$ is
the center of mass energy of the $e^{+} e^{-}$ system, $t = M_{L}^{2}
- \frac{s}{2} (1 - \beta \cos \theta)$ and {} $u = M_{L}^{2} - \frac{s}{2}
(1 + \beta \cos \theta)$, where $\theta$ is the angle between the exotic
lepton and the incident electron, in the c.m. frame.

As for the NLC and for the CLIC, the contribution of the $Z$ exchange
is expected to be smaller than the photon exchange, since the couplings of exotic leptons to $Z$ have a weak interaction strength and the mixing angles are also weak, while for the photon we have a
electromagnetic strength only.

\hskip 0.5cm

The production of exotic quarks can be studied through the analysis of
the process $e^{+} e^{-} \rightarrow \bar{Q} Q$, involving also the
exchange of a photon and a boson $Z$ in the $s$ channel. Using the
interactions Lagrangians, Eqs. ($1$) and ($2$), we obtain for the cross section the expression

\begin{eqnarray} 
\left (\frac{d \sigma}{d\cos \theta} \right )_{\bar{Q} Q} =&&\frac{\beta_{Q} \alpha^{2} \pi e_{q}^{2}}{s^{3}} \Biggl (2 s M^{2}_{Q} + (M^{2}_{Q} - t)^{2} + (M^{2}_{Q} - u)^{2} \Biggr ) \nonumber \\
&&- \frac{\beta_{Q} \pi  \alpha^{2} eq}{2 \sin^{2}\theta_{W} \cos^{2}\theta_{W} (s - M^{2}_{Z}) s^{2}} \Biggl (2 s M^{2}_{Q} g_{V}^{QQ} g_{V}^{l} + (M^{2}_{Q} - t)^{2} \nonumber \\
&&(g_{V}^{QQ} g_{V}^{l} -g_{A}^{QQ} g_{A}^{l} ) + (M^{2}_{Q} - u)^{2} (g_{V}^{QQ} g_{V}^{l} + g_{A}^{QQ} g_{A}^{l} )  \Biggr ) \nonumber \\
&&+\frac{\beta_{Q} G_{F}^{2} M_{Z}^{4}}{8 \pi s} \Biggl [\frac{1}{s - M_{Z}^{2}}  \Biggl ((g_{V}^{{ij}^{2}} + g_{A}^{{ij}^{2}}) (g_{V}^{{e}^{2}} + g_{A}^{{e}^{2}}) ((M_{L}^{2} - u)^{2}    \nonumber  \\
&&+(M_{L}^{2} - t)^{2}) + 2 (g_{V}^{{ij}^{2}} - g_{A}^{{ij}^{2}}) (g_{V}^{{e}^{2}} + g_{A}^{{e}^{2}}) s M_{L}^{2}    \nonumber  \\
&&+ 4 g_{V}^{ij} g_{A}^{ij} g_{V}^{e} g_{A}^{e} 
 ((M_{L}^{2} - u)^{2} - (M_{L}^{2} - t)^{2} ) \Biggr ) \Biggr ]    \;,
\end{eqnarray}
where $\beta_{Q} = \sqrt{1- 4 M_{Q}^{2}/s}$ is the velocity of
exotic-quark in the c.m. of the process, $e_{q}$ is the charge of the quark, $Q$ are the quark and $\bar{Q}$ the antiquark, respectively. We  have for this production process the same high energy behavior as for the exotic leptons.

Finally, for the process $e^{+} e^{-} \rightarrow L^{0} L^{0}$, we have
the following result, involving a charged current,

\begin{eqnarray} 
\left (\frac{d \sigma}{d\cos \theta} \right )_{L^{0} L^{0}} =&&\frac{\beta_{L^{0}} G_{F}^{2} M_{W}^{4}}{32 \pi s (t - M_{W}^{2})^{2}}  \Biggl (\frac{s^{2}}{2}(1 + \beta_{L^{0}}^{2}) (C_{V}^{{eL^{0}}^{2}} - C_{A}^{{eL^{0}}^{2}})^{2}     \nonumber  \\
&&+ (M_{L^{0}}^{2} - u)^{2} (C_{V}^{{eL^{0}}^{2}} + C_{A}^{{eL^{0}}^{2}})^{2} + 4 (M_{L^{0}}^{2} - u)^{2} C_{V}^{{eL^{0}}^{2}} C_{A}^{{eL^{0}}^{2}} \Biggr )         \;, 
\end{eqnarray}
where $\beta_{L^{0}} = \sqrt{1- 4 M_{L^{0}}^{2}/s}$ is the velocity of
exotic-neutrino in the c.m. of the process, $M_{W}$ is the mass of
the boson $W$ and $L_{0}$ is the neutrino.


\section{RESULTS AND CONCLUSIONS}

In Figs. $1$ and $2$ we present the cross section for the process
$e^{+} e^{-} \rightarrow L^{+} L^{-}$, involving the three models studied
here: the VSM, VDM and FMF models, for the NLC and CLIC. In all
calculations we take $\sin^{2} {\theta_W} = 0.2315$, $M_Z = 91.188$ GeV and $M_{W} = 80.33$ GeV.

Considering that the only limitation for the discovery of charged
heavy leptons is the total number of events \cite{bar,hin}, we take
into account that the expected integrated luminosity for the NLC will
be of order of $6.10^{4} pb^{-1}/yr$ which gives a total of: $\simeq
2.10^{4}$ lepton pairs produced per year, for the FMF model, and 
$\simeq 10^{4}$ for both the VSM and VDM models. We note, therefore, that it will be very difficult to distinguish these two class of leptons, whereas for fermion mirror fermion leptons the difference away from threshold is $\simeq 20 \%$ and consequently it will be easy to separate the FMF leptons from the other two classes of particles. Now, considering that the integrated luminosity for the CLIC
will be of order of $2.10^{5} pb^{-1}/yr$, then the statistics that we
can expect for this collider is somewhat richer, although the cross
sections is for $1$ TeV smaller. Then the total number of lepton pairs produced per year for this machine will be $\simeq 32.10^{3}$ for the FMF model and $\simeq 22X10^{3}$ for both the VSM and VDM models (here it is valid the same argumentation that we have above for the NLC, in respect to distinguish the models). We also note  that the production cross section for these exotic leptons are larger than the production of single exotic leptons \cite{sim1}.

The decay modes of charged exotic heavy leptons depend on the mass difference, $\Delta = M_{L} - M_{L^{0}}$, where $M_{L^{0}}$ is the heavy neutrino mass . If $\Delta > M_{W}$, the main decay mode of the heavy charged lepton is into a real $W$ and $L^{0}$, since this
neutrino will go undetected. The signal for heavy lepton production is
$W^{+} W^{-}$, plus missing transverse momentum.

On the order hand, requiring that the $W$'s decay leptonically, the main signal for $L^{+} L^{-}$ pair is $\ell^{+} \ell^{-}$ plus missing transverse momentum ($\ell=\mu, e$). The processes that have the same signature of this signal are $e^{+} e^{-} \rightarrow \tau^{+} \tau^{-}, W^{+} W^{-}$ and $Z^{0} Z^{0}$. These backgrounds can be eliminated by calculating the invariant mass of the charged fermion pair. Further on, the signal is very striking since it consists of a pair of
$\ell^{+} \ell^{-}$, with approximately the same invariant mass.

In Fig. $3$ we compare the cross section for $e^{+} e^{-}
\rightarrow L^{+} L^{-}$ for FMF (which is the largest), where we
take for the mass of exotic leptons $M_{L} = 200$ GeV, along with the
production cross sections for $W^{+} W^{-}$, $Z^{0} Z^{0}$ and $\tau^{+} \tau^{-}$. We see from these results that the production of $W^{+} W^{-}$ is larger than in $L^{+} L^{-}$ and the production of
$Z^{0} Z^{0}$ and $\tau^{+} \tau^{-}$ are smaller than in $L^{+} L^{-}$.  Here we must do a careful analysis of Monte Carlo to separate
the signal from the background.

Fig. $4$ shows the distribution cross section for the production of heavy lepton $L^{-}$ of mass equal to $200$ GeV and c.m. energy,  $\sqrt{s}$, equal to $500$ GeV. We can observe from this result that this distribution is quite similar for both the VSM and the VDM and it  gives a closely symmetrical distribution, while for the FMF it gives a peaked distribution for $\cos \theta \simeq 1$.

In Figs. $5$ and $6$ we show the cross sections for the production of
exotic quarks $e^{+} e^{-} \rightarrow \bar{Q} Q$, for the VSM, VDM and FMF models, in the colliders NLC and CLIC. We see from these
results that we can expect for the first collider a total of $\simeq
14.10^{3}$ heavy quark pairs produced per year for the VDM, while, for both the FMF and VSM models, we estimate a total of $\simeq 9.10^{3}$ events per year. We note that it will not be easy to distinguish the FMF and VSM - quarks, while for the VDM - leptons, which are away from threshold, gives $\propto 15 \%$ more events than the other two  models and, consequently, it will not be difficult to separate the FMF and VSM from the VDM - quarks. These analysis are similar for the CLIC. For the second collider we expect a total of $\simeq 10^{4}$ quarks for both the VSM and FMF models, while, for the VDM model, we have a total of $\simeq 14.10^{3}$ events per year.  We obtain, then, more events for the VDM model. This is due to the coupling constants, which in the case of VDM, are larger than the ones for the FMF and  VSM models. We have considered here only the heavy quark U, since for the quark D the cross section is of the same order. We obtain,  therefore, that the production cross-sections for these exotic quarks are competitive, when compared with the ones for the production of single exotic quarks \cite{sim3}.

In Fig. $7$ we compare the cross section for $e^{+} e^{-} \rightarrow  \bar{Q} Q$ for the VDM model (whose cross-section is the largest),  with the production cross section for $\bar{t} t$.  We see from these results that the production of $\bar{t} t$ is larger than that for the  production for $\bar{U} U$ and $\bar{D} D$. Again, to separate the signal of the background, a careful analysis of Monte Carlo must be done.

Fig. $8$ shows the distribution of cross-section for the production  of a heavy quark U, with mass equal to $200$ GeV and c.m. energy,  
$\sqrt{s}$, equal to $500$ GeV. We observe that the distribution for  the VSM and VDM models are symmetrical one to another, while for the FMF it gives a peaked distribution for $\cos \theta \simeq -1$. We also note that the quark D have the same behavior as the
quark U.

In Figs. $9$ and $10$ we show the cross section for the production of
exotic neutrinos, $e^{+} e^{-} \rightarrow \bar{L^0} L^0$, for the VSM, VDM and FMF models, in the colliders NLC and CLIC. We see from
these results that we can expect, for both the VSM and the VDM models, in the NLC, a total of  around $100$ heavy neutrinos pairs produced per year. The number of events are equal because we have taken for the mixing angles the same values. For the FMF model, we obtain a total of $\simeq 300$ events per year. 
We also have that, in the CLIC, for both the VSM and VDM models, it can be produced a total of $\simeq 200$ pairs of exotic neutrinos and for the FMF model, it can be observed a total of $900$ events. The number of events here are very small in comparison with the production of exotic leptons and quarks, because the $\gamma$ and $Z$ take no part in the production of exotic-neutrinos. Therefore these productions are not competitive with the production of single exotic neutrinos \cite{sim1}.

\acknowledgments

I would like to thank the Departamento de F\'{\i}sica Matem\'atica
(USP) for its kind hospitality, where part of this work was done, to Prof. O. J. P. \'Eboli for calling the attention to some points and to Prof. R. O. Ramos for a careful reading of the manuscript.

\newpage



\newpage

\begin{center}
FIGURE CAPTIONS
\end{center}

\vspace{0.5cm}

{\bf Figure 1}: Total cross section for the process $e^{+} e^{-} 
\rightarrow L^{+} L^{-}$ as a function of $M_{L}$ at $s = 500$ GeV: 
(a) vector singlet model (dotted line); (b) vector doublet model
(dashed line); (c) fermion mirror fermion (solid line).

{\bf Figure 2}: Total cross section for the process $e^{+} e^{-} \rightarrow 
L^{+} L^{-}$ as a function of $M_{L}$ at $s = 1000$ GeV: (a) vector
singlet model (dotted line); (b) vector doublet model (dashed line);
(c) fermion mirror fermion (solid line).

{\bf Figure 3}: Total cross section versus the total c.m. energy $\sqrt{s}$ for the 
following processes (a) $e^{+} e^{-} \rightarrow L^{+} L^{-}$ for the
case of FMF (dotted line); (b) $e^{+} e^{-} \rightarrow W^{+} W^{-}$
(dashed line); (c) $e^{+} e^{-} \rightarrow Z Z$ (solid line); (d) $
e^{+} e^{-} \rightarrow \tau^{+} \tau^{-}$ (long dashed line).

{\bf Figure 4}: Angular distribution of the primary lepton of the neutral 
current for the (a) vector singlet model (dotted line); (b) vector
doublet model (dashed line); (c) fermion mirror fermion (solid line).

{\bf Figure 5}: Total cross section for the process 
$e^{+} e^{-} \rightarrow \bar{Q} Q$ as a function of $M_{L}$ at $s =
500$ GeV: (a) vector singlet model (dotted line); (b) vector doublet
model (dashed line); (c) fermion mirror fermion (solid line).

{\bf Figure 6}: Total cross section for the process $e^{+} e^{-} \rightarrow 
\bar{Q} Q$ as a function of $M_{L}$ at $s = 1000$ GeV: 
(a) vector singlet model (dotted line); (b) vector doublet model
(dashed line); (c) fermion mirror fermion (solid line).

{\bf Figure 7}: Total cross section versus the total c.m. energy $\sqrt{s}$ for the 
following processes (a) $e^{+} e^{-} \rightarrow \bar{U} U$ for the case
of VDM (dotted line); (b) $e^{+} e^{-} \rightarrow \bar{D} D$ for the
case of VDM (dashed line); (c) $e^{+} e^{-} \rightarrow \bar{t} t$ (solid line).

{\bf Figure 8}: Angular distribution of the primary quark of the neutral current for
the (a) vector singlet model (dotted line); (b) vector doublet model
(dashed line); (c) fermion mirror fermion (solid line).

{\bf Figure 9}: Total cross section for the process $e^{+} e^{-} \rightarrow N N $ 
as a function of $M_{N}$ at $s = 500$ GeV: (a) vector singlet model
(dotted line); (b) vector doublet model (dashed line); (c) fermion
mirror fermion (solid line).

{\bf Figure 10}:  Total cross section for the process $e^{+} e^{-} \rightarrow N N$ 
as a function of $M_{N}$ at $s = 1000$ GeV: (a) vector singlet model
(dotted line); (b) vector doublet model (dashed line); (c) fermion
mirror fermion (solid line).

\newpage

\begin{figure}[b]
\epsfysize=18cm 
{\centerline{\epsfbox{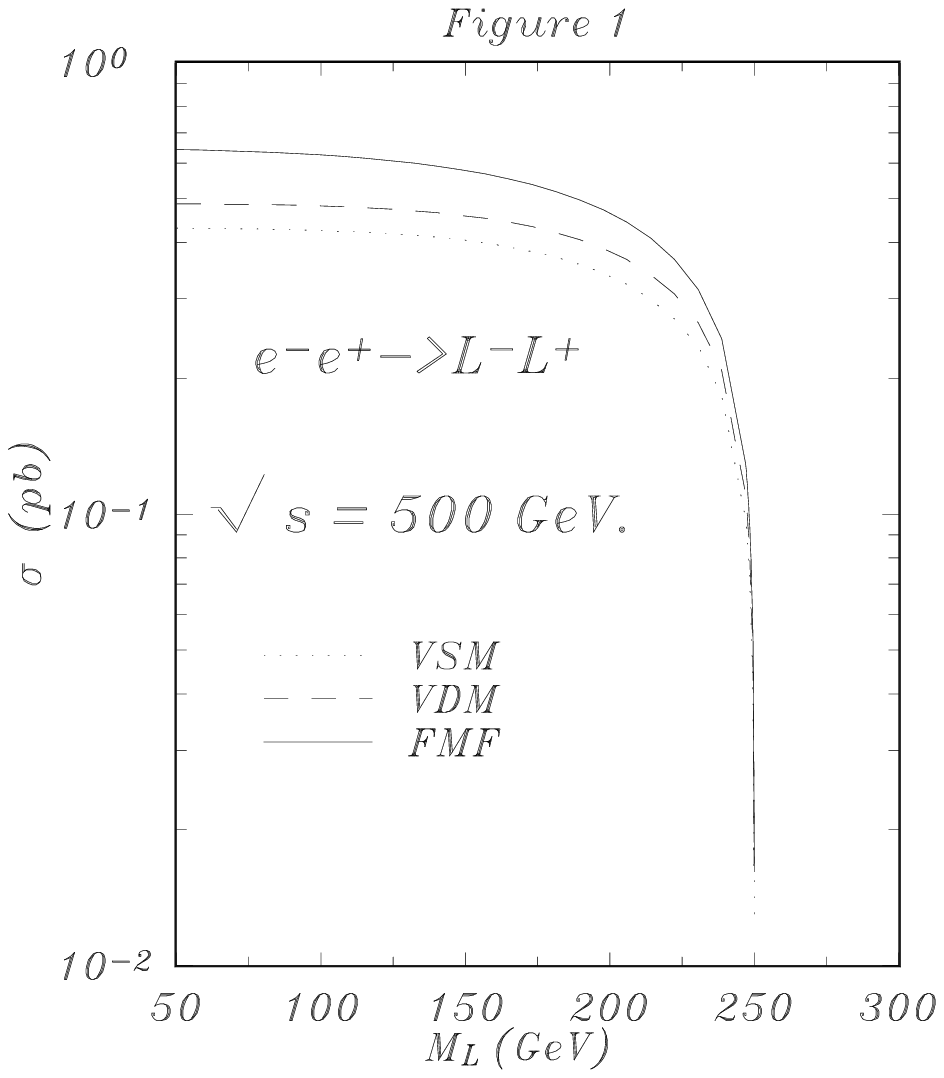}}}

\vspace{1cm}

\end{figure}

\newpage

\begin{figure}[b]
\epsfysize=18cm 
{\centerline{\epsfbox{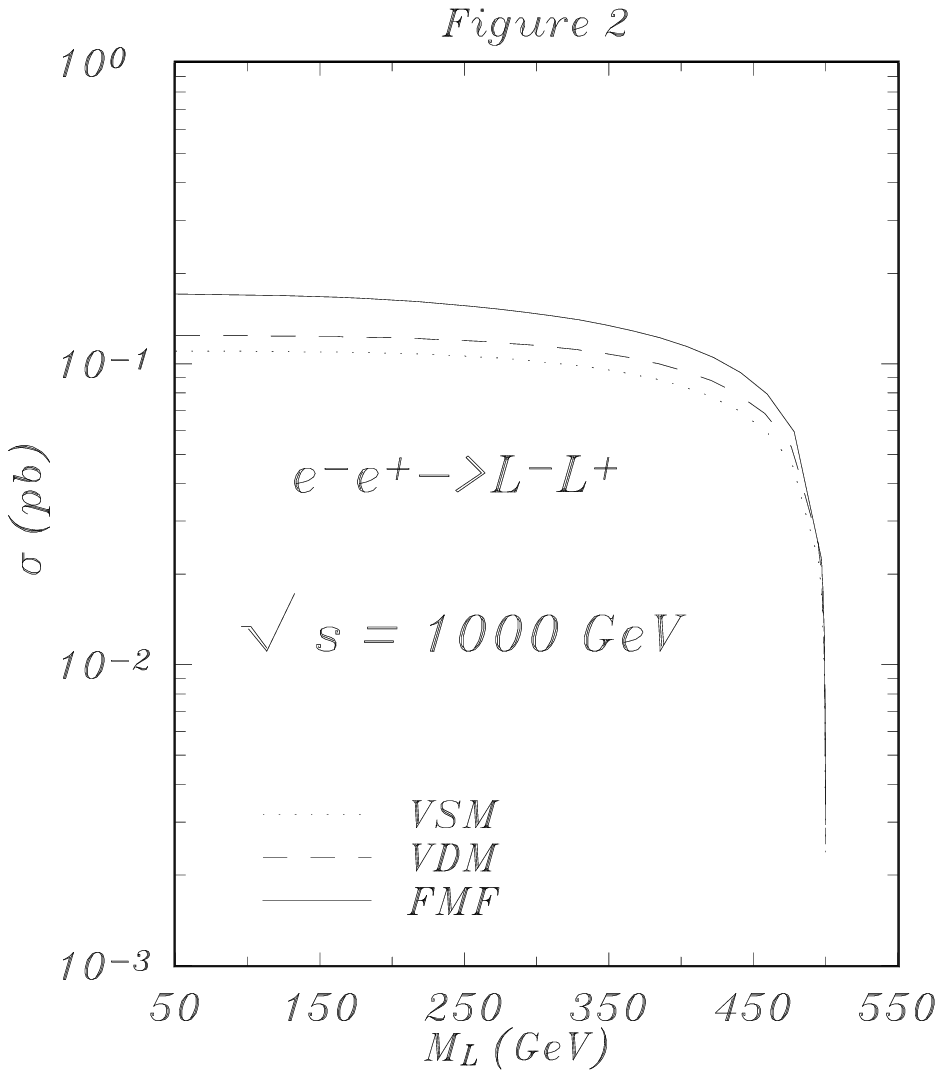}}}

\vspace{1cm}

\end{figure}

\newpage

\begin{figure}[b]
\epsfysize=18cm 
{\centerline{\epsfbox{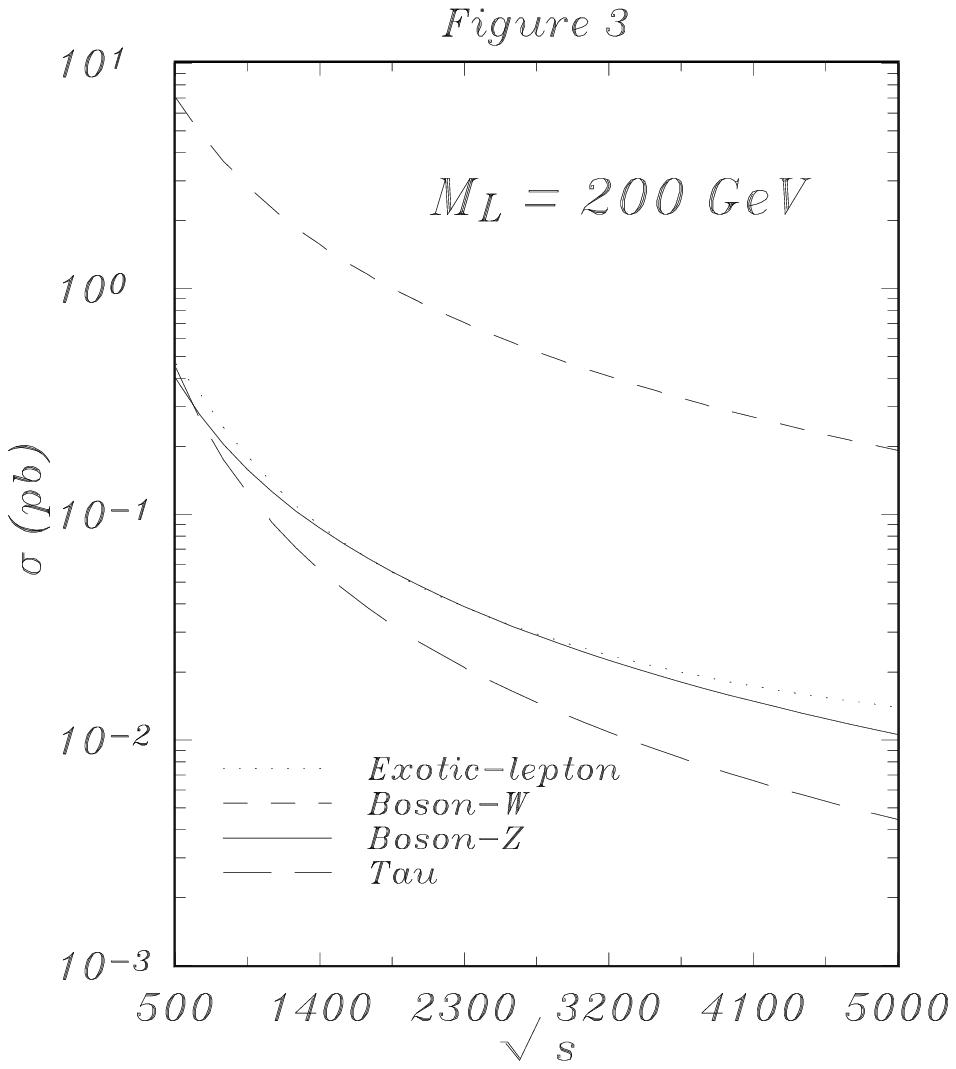}}}

\vspace{1cm}

\end{figure}

\newpage

\begin{figure}[b]
\epsfysize=18cm 
{\centerline{\epsfbox{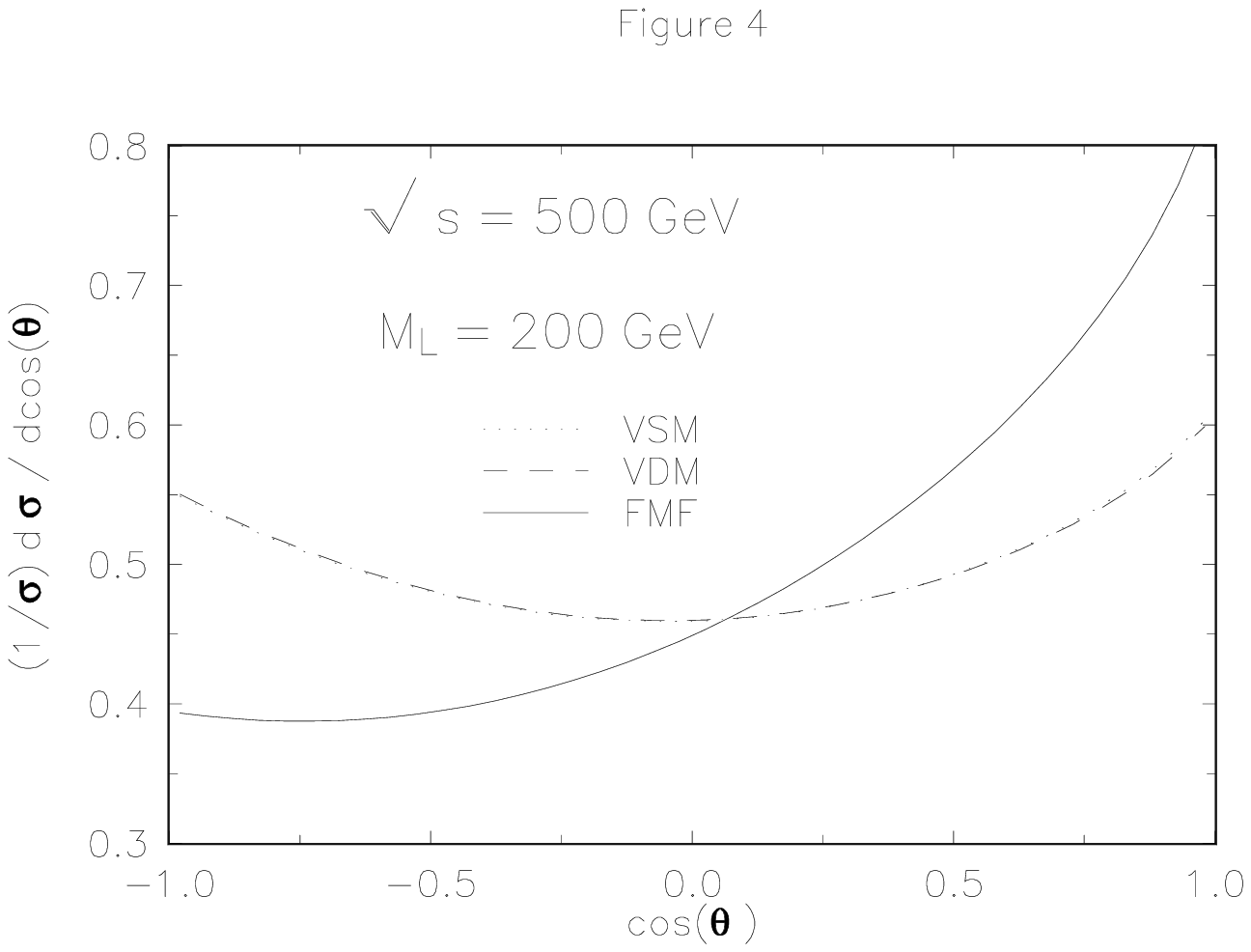}}}

\vspace{1cm}

\end{figure}

\newpage

\begin{figure}[b]
\epsfysize=18cm 
{\centerline{\epsfbox{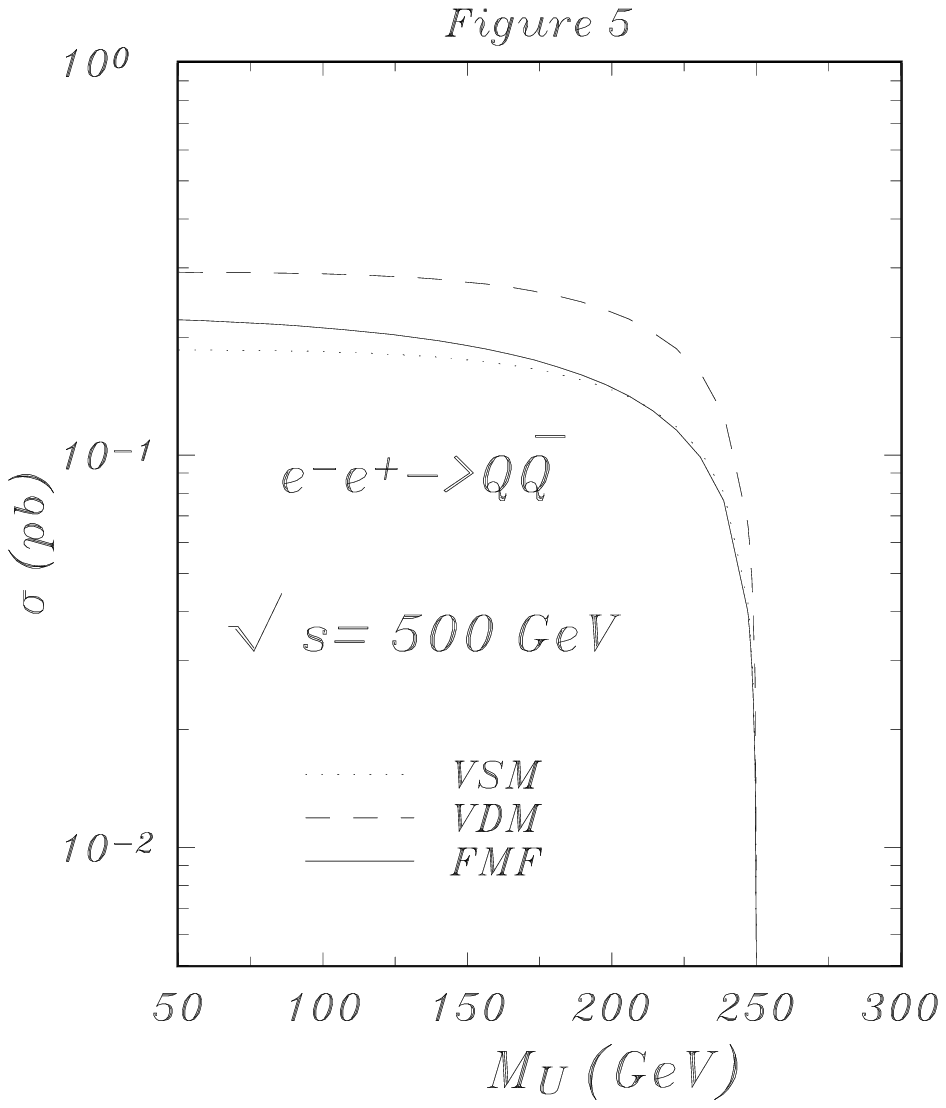}}}

\vspace{1cm}

\end{figure}

\newpage

\begin{figure}[b]
\epsfysize=18cm 
{\centerline{\epsfbox{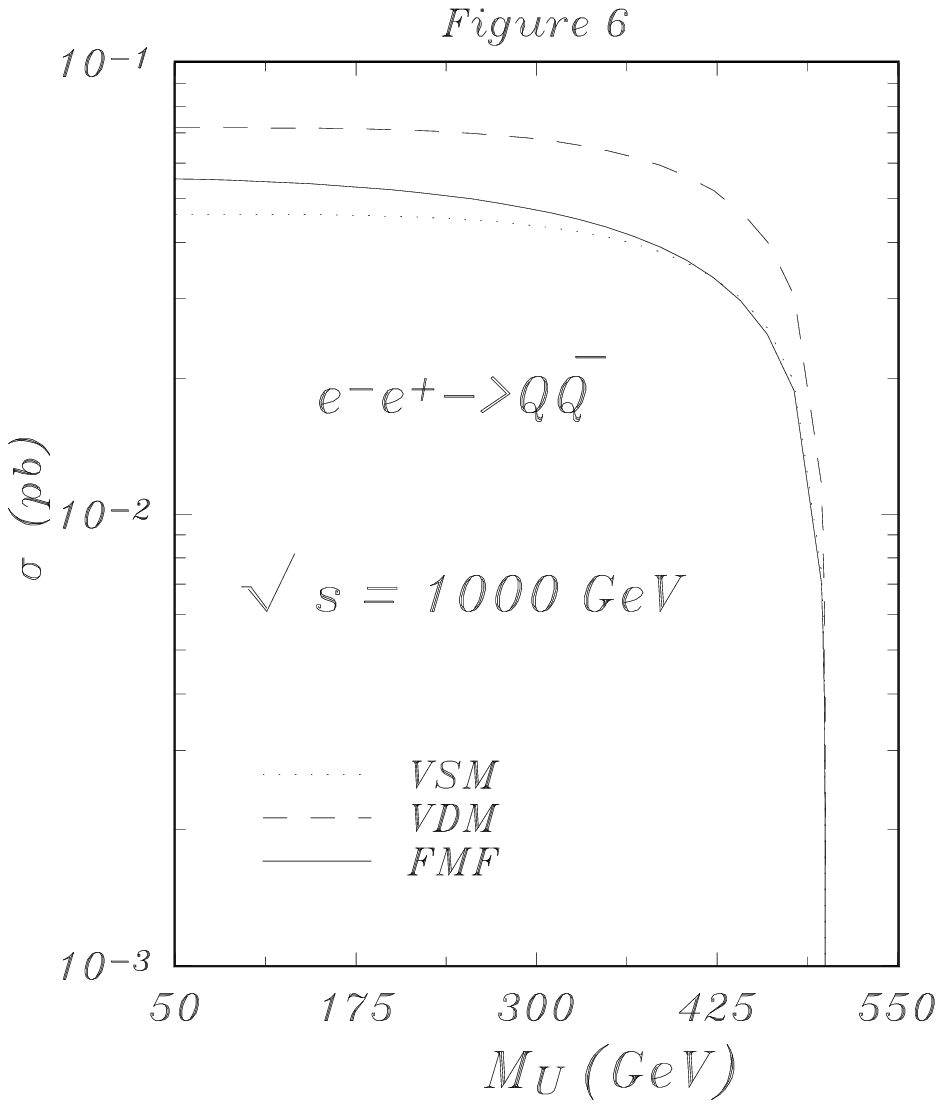}}}

\vspace{1cm}

\end{figure}

\newpage

\begin{figure}[b]
\epsfysize=18cm 
{\centerline{\epsfbox{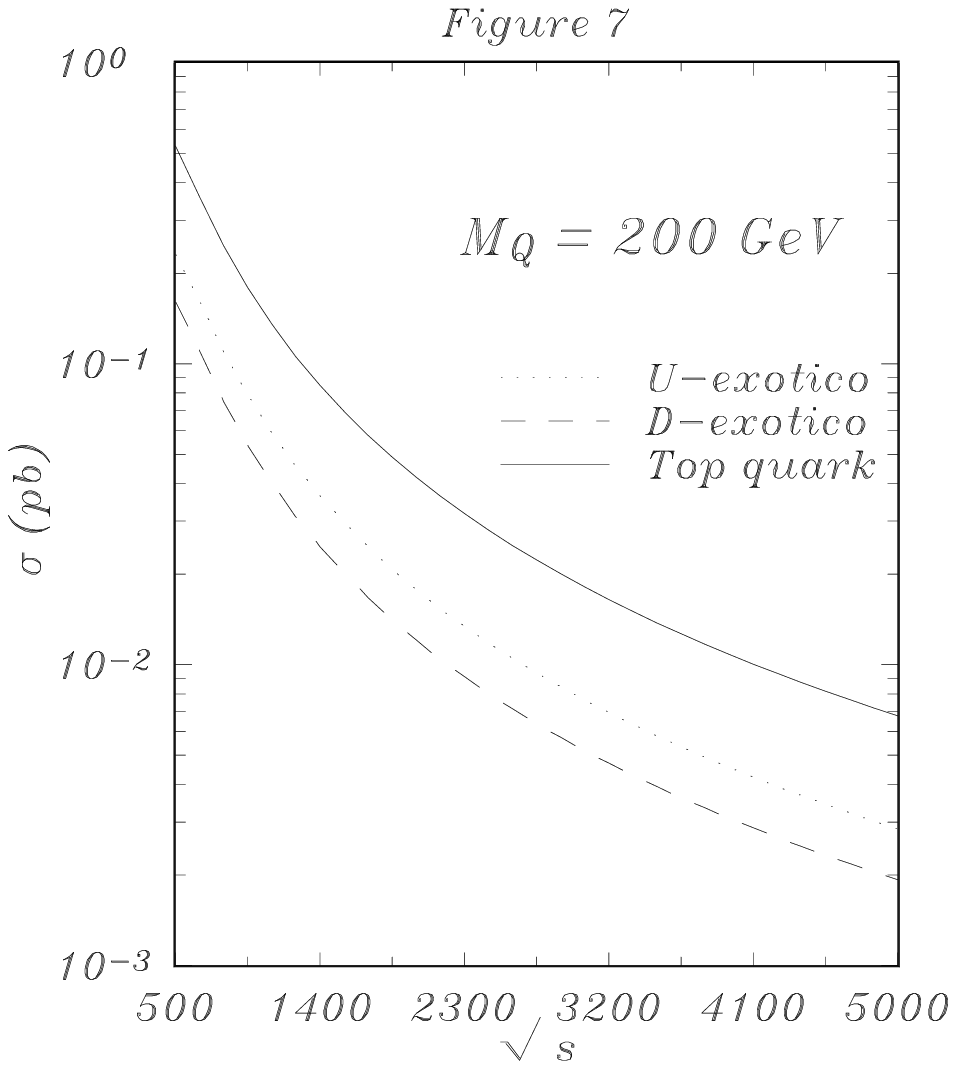}}}

\vspace{1cm}

\end{figure}

\newpage

\begin{figure}[b]
\epsfysize=18cm 
{\centerline{\epsfbox{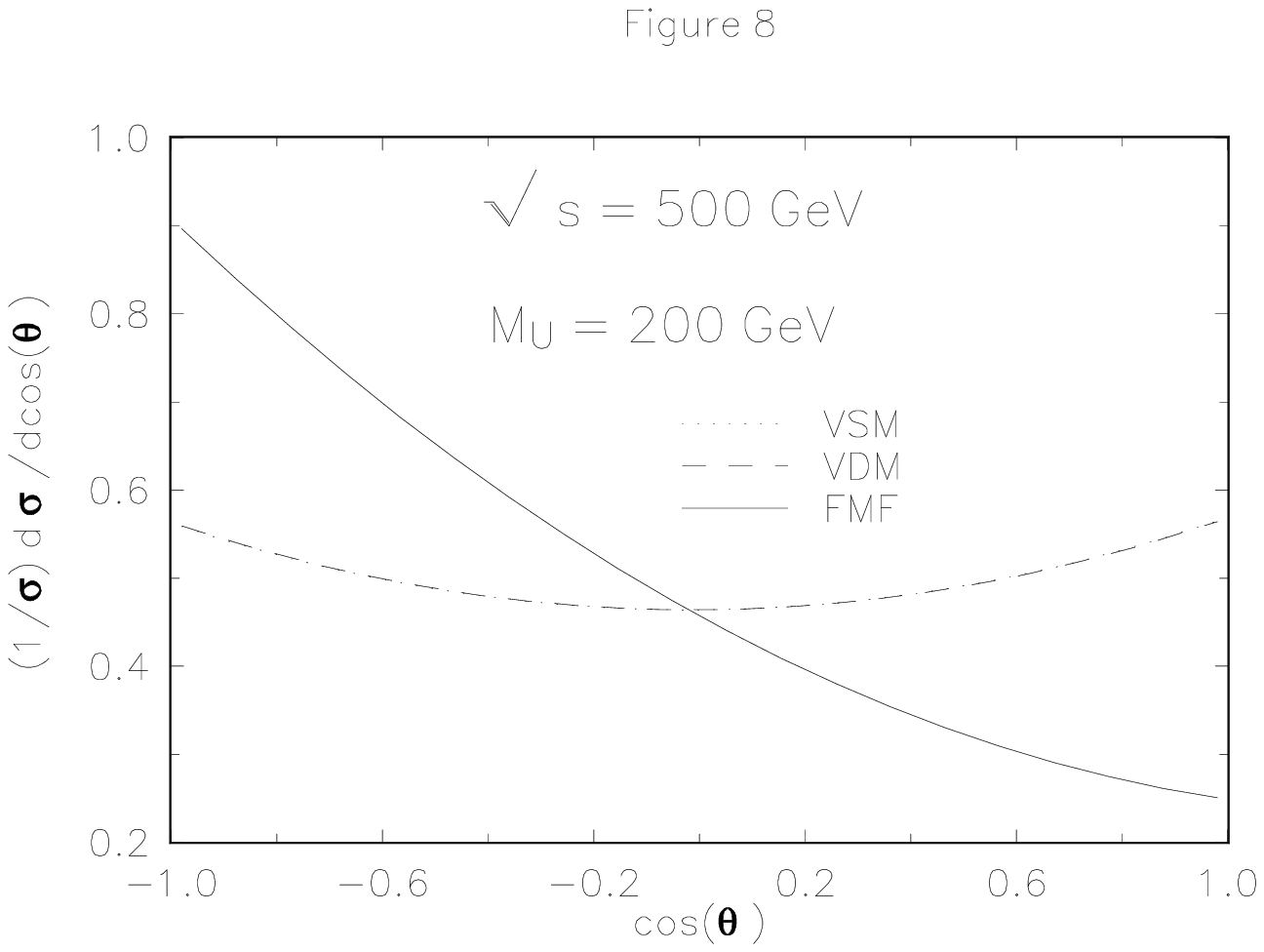}}}

\vspace{1cm}

\end{figure}

\newpage

\begin{figure}[b]
\epsfysize=18cm 
{\centerline{\epsfbox{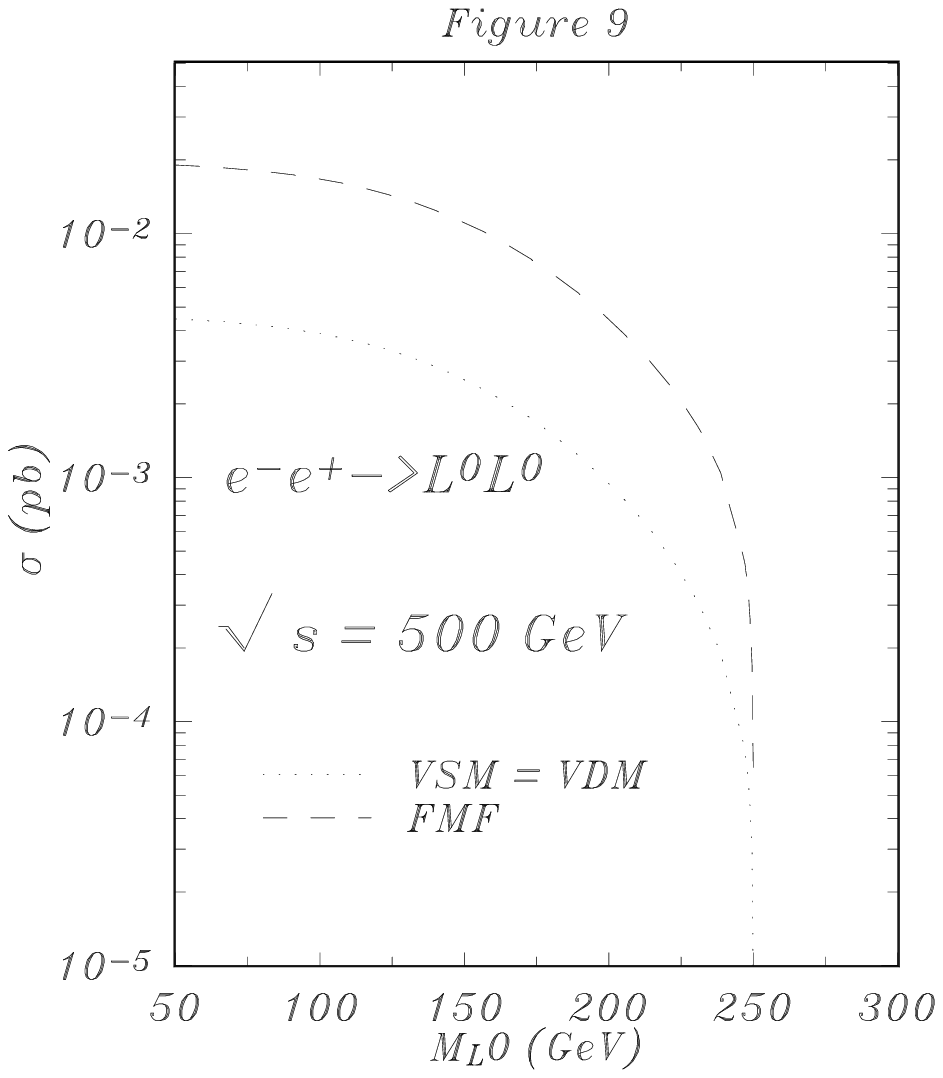}}}

\vspace{1cm}

\end{figure}

\newpage

\begin{figure}[b]
\epsfysize=18cm 
{\centerline{\epsfbox{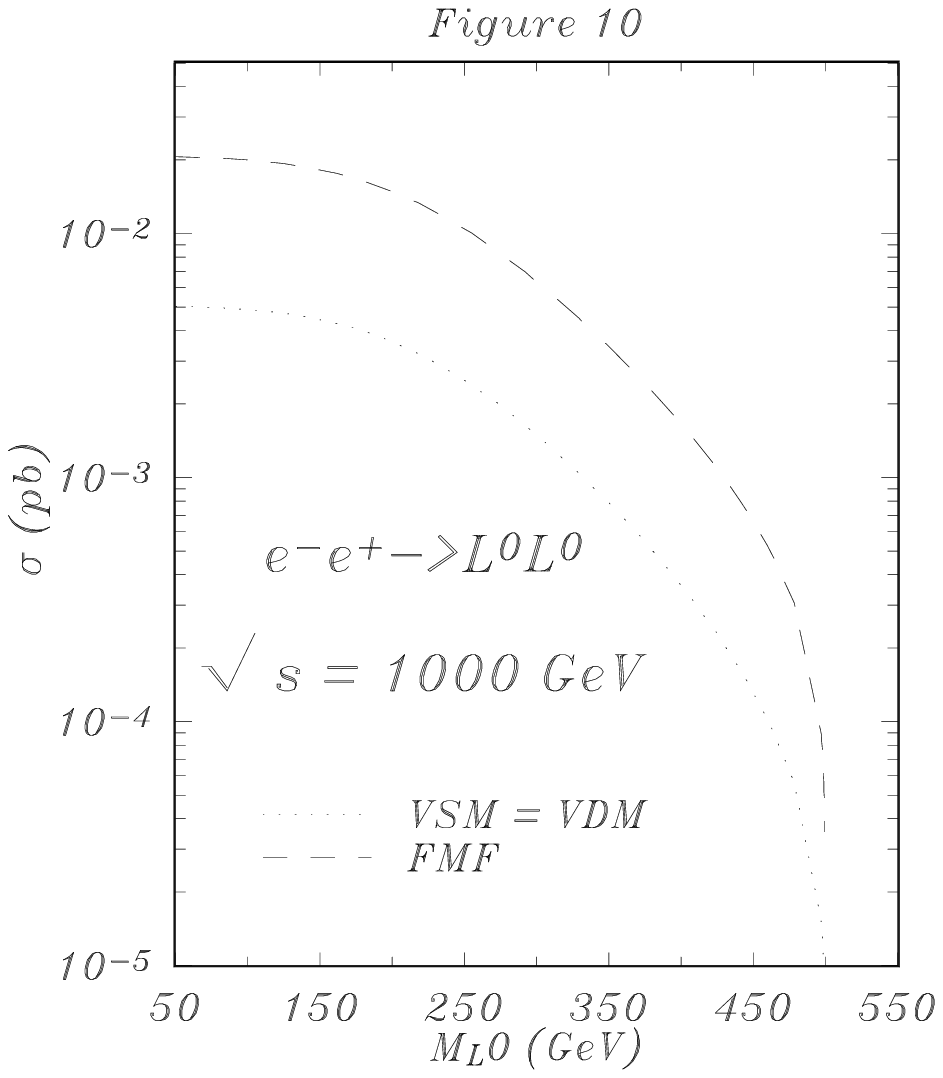}}}

\vspace{1cm}

\end{figure}

\end{document}